\documentclass[11pt]{article}
\usepackage{moriond,cite}
\usepackage[dvips]{graphicx}
\graphicspath{{./figures/}}

\usepackage{units}

\begin{document}
\vspace*{4cm}
\title{CURRENT NUMI/MINOS OSCILLATION RESULTS}

\author{ A. HABIG,\\ for the MINOS Collaboration }

\address{Department of Physics, University of Minnesota Duluth\\
10 University Dr., Duluth, MN 55812 USA}

\maketitle\abstracts{
The MINOS experiment is now making precise measurements of the $\nu_\mu$
disappearance oscillations seen in atmospheric neutrinos, and will
extend our reach towards the so far unseen $\theta_{13}$ by looking for
$\nu_e$ appearance in the $\nu_\mu$ beam.  It does so by using the
intense, well-understood NuMI neutrino beam created at Fermilab and
observing it \unit{735}{km} away at the Soudan Mine in Northeast Minnesota.
Results from MINOS' first two years of operations will be presented.
}

\section{Introduction}\label{sec:intro} 

Results from the Super-Kamiokande experiment used neutrinos produced by
cosmic ray interactions with the upper atmosphere to show that muon
neutrinos ($\nu_\mu$) of energies from a few hundred MeV through TeV
oscillate to tau neutrinos ($\nu_\tau$) as they travel the tens to
thousands of kilometers through the earth to the
detector\cite{superkprd}.  This implies that neutrinos have mass, a
finding of fundamental importance to both particle physics and
astrophysics. The K2K experiment used a beam of neutrinos shot across
Japan to the Super-K detector to confirm this result in a controlled
fashion\cite{k2k-prd}.  The MINOS (\underline{M}ain \underline{I}njector
\underline{N}eutrino \underline{O}scillation \underline{S}earch)
experiment has unambiguously confirmed this result.  MINOS will
precisely measure the oscillation parameters using the intense,
well-calibrated NuMI (\underline{Neu}trinos at the \underline{M}ain
\underline{I}njector) beam of neutrinos generated at Fermilab.  This
neutrino beam was commissioned in early 2005 and is aimed toward the
Soudan Underground Physics Laboratory in northeastern Minnesota.  The
neutrinos are observed by similar magnetized steel/scintillator
calorimeters near their origin in Fermilab and after traveling
\unit[735]{km} to Soudan.

Differences in signals between the two detectors have already provided
the best measurement yet of $\nu_\mu\leftrightarrow\nu_\tau$ flavor
oscillations in a long-baseline accelerator experiment, using the first
two years operation of the NuMI neutrino beam\cite{minosprd}.  With
more data, MINOS will reach its projected sensitivity to this mixing,
improved sensitivity to any sub-dominant $\nu_e$ modes (a probe of
$\theta_{13}$) and high statistics neutrino cross section studies.  This
paper presents the current result on $\nu_\mu\leftrightarrow\nu_\tau$
oscillations, the first look at the spectrum of neutral current (``NC'')
events in the MINOS near detector, the methods which will be used to
search for $\nu_e$ appearance, and new data-driven sensitivities to
$\theta_{13}$.

\subsection{The NuMI Beam}\label{sec:numi}

The NuMI neutrino beam\cite{Abramov:2001nr} uses \unit[120]{GeV} protons
from the Main Injector synchrotron at Fermilab incident upon a graphite
target.  90\% of the primary protons interact over the two
interaction-length long target, producing showers of $\pi$ and $K$
mesons.  These showers are focused by a pair of parabolic aluminum
``horns'', pulsed electromagnets carrying current sheaths which focus
the mesons into a beam.  This beam is sent down a \unit[1]{m} radius,
\unit[675]{m} long decay pipe.  While in this pipe the mesons have a
chance to decay into muons and muon neutrinos, but few of the muons have
enough time to further decay before they are absorbed at the end of the
pipe, a decay which would produce electron and anti-muon neutrinos.  The
resulting neutrino beam is thus composed of approximately
92.9\%~$\nu_\mu$, 5.8\%~$\overline{\nu}_{\mu}$, 1.2\%~$\nu_e$ and
0.1\%~$\overline{\nu}_{e}$ for the low-energy (``LE'') beam configuration.

\begin{figure}[tb]
  \begin{center}
    \includegraphics[width=0.9\textwidth]{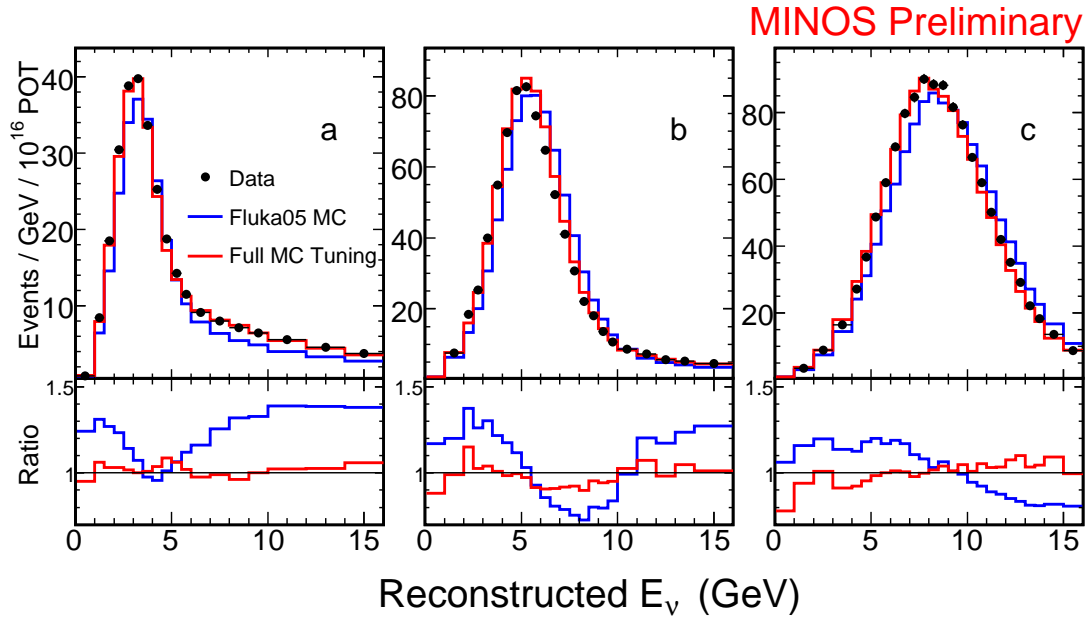}
    \caption{The measured energy spectrum of neutrinos from the NuMI beam
      observed by the MINOS near detector (top) and the ratios of data
      and expectations (bottom).  Data points are the black dots, the
      untuned MC predictions are the blue curves, and MC predictions
      after tuning on hadronic $x_F$ and $p_t$ simultaneously across
      many different beam configurations are the red curves.}
    \label{fig:beamspectra}
  \end{center}
\end{figure}

The target and horns are movable with respect to each other, allowing
different focusing optics.  The result is a beam which is configurable
in energy, as seen in Fig.~\ref{fig:beamspectra}.  The LE configuration
produces a spectral peak closest to the first oscillation minima, given
the oscillation parameters measured by Super-K and the \unit[735]{km}
baseline to the far detector.  Moving the target with respect to the
horns produces the ``pME'' and ``pHE'' beams peaked at medium and higher
energies.  While not at ideal energies for the $\nu_\mu$ disappearance
analysis, these beams are much more intense ($\sim$970 and 1340 neutrino
events at the far detector per 10$^{20}$ protons on target, compared to
$\sim$390 for the LE beam) and provide extra handles when using the near
detector data to model the beam's properties.  The MINOS near detector
is only a km away from the target, so even the LE beam produces around
10$^7$ neutrino interactions per 10$^{20}$ protons on target, a very
high statistics sample of this weakly interacting particle.  The beam
currently delivers 3.1$\times$10$^{12}$ protons over a \unit[12]{$\mu$s}
spill every \unit[2.2]{s} for an average power of \unit[270]{kW}.  The
NuMI beam has been operational since march of 2005, and to date (of this
conference, March 2008) has delivered more than 4$\times$10$^{20}$
protons on target.

\subsection{The MINOS Detectors}\label{sec:minos}

The MINOS experiment observes the NuMI beam with two detectors, ``near''
and ``far''.  A third ``calibration'' detector was exposed to beams of
protons, pions, electrons and muons from the the CERN
PS~\cite{Adamson:2005cd} to determine detector response.  The near
detector at Fermilab is used to characterize the neutrino beam with high
statistics and is \unit[1]{km} downstream from the NuMI target.
The far detector is an additional \unit[734]{km} downstream.  This
experiment compares the spectra of different types of neutrino
interactions at these two detectors to test oscillation
hypotheses.

All three MINOS detectors are steel-scintillator sampling
calorimeters\cite{minosnim} made of alternate planes of
\unit[4.1$\times$1]{cm} cross section plastic scintillator strips and
\unit[2.54]{cm} thick steel plates.  The near and far detectors have
magnetized steel planes.  The calibration detector was not magnetized as
the incoming particle momenta were known.  The extruded polystyrene
scintillator strips are read out with wavelength-shifting fibers and
multi-anode photomultiplier tubes.  The far detector is \unit[705]{m}
underground in Soudan, MN, in a disused iron mine currently operated as
a State Park by the Minnesota Department of Natural Resources.  The
\unit[5,400]{metric ton} far detector consists of 486 \unit[8]{m}-wide
octagonal steel planes interleaved with planes of plastic scintillator
strips.

The \unit[282]{plane}, \unit[980]{metric ton} MINOS near detector is
located at the end of the NuMI beam facility at Fermilab in a
\unit[100]{m} deep underground cavern.  While the NuMI beam has diverged
to a mile wide at Soudan, at the near detector it is mostly contained in
a meter-wide area, allowing a smaller detector and a factor of 10$^6$
higher neutrino rate.

The much smaller calibration detector was used to measure the detailed
responses of the MINOS detectors in a charged-particle test beam. This
\unit[12]{ton} detector consisted of \unit[60]{planes} of unmagnetized
steel and scintillator, each 1$\times$1 m$^2$\cite{Adamson:2005cd}.  It
measured the energy and topological responses expected in the the near
and far detectors, including the different electronics used in both
larger devices.  The energy responses of the three MINOS detectors were
normalized to each other by calibrating with cosmic-ray muons.

\section{Data Analysis}\label{sec:analysis}

MINOS beam-based data is analyzed using a 
``blind analysis''.  This method avoids looking at the actual data
containing the physics being studied until the very end, removing
potential biases and increasing confidence in the final result.  Monte
Carlo (``MC'') predictions are tuned and verified using data not
sensitive to the physics in question ({\it e.g.} near detector data
which is at too short a baseline to have experienced oscillations), and
analysis cuts and techniques developed solely using simulated data.
Only after these techniques are optimized and set are the sensitive data
(in this example, the far detector oscillated data) revealed.  All three
of the results discussed in this paper are blind analyses, and are at
different stages in the process.

The first step, common to all beam-based analyses, is to understand the
beam itself.  A detailed MC tracks simulated particles through the
proton-meson-neutrino chain described in Sec.~\ref{sec:numi}, to create
an expected neutrino spectrum at the near detector.  This MC is
developed and crosschecked with information from the NuMI beam
monitoring system, including a hadron monitor in the absorber at the end
of the decay pipe and three muon monitors further downstream.  As can be
seen in the blue curves in Fig.~\ref{fig:beamspectra}, this does a
decent but not perfect job of predicting the observed neutrino spectra
in the near detector.  Further tuning is done by reweighting hadronic
$x_F$ and $p_t$ in the MC simultaneously across seven different beam
configurations and comparing to real near detector data, as the hadronic
models have the most theoretical uncertainty.  Four additional
beam configurations (with different horn focusing currents) beyond those
shown are included in this fit, and the resulting tuned predictions are
the red line in Fig.~\ref{fig:beamspectra}.  With the MC truth
information in hand, a far detector prediction can be made by applying
changes due to mundane things like geometrical and kinematic factors, or
more exciting things like neutrino oscillations.

With a beam MC prediction in hand, topological features in the near
detector data can be examined.  Fitters to find tracks and neutrino
interaction vertices, shower-finding algorithms, and particle
identification (``PID'') routines can be developed, tested, and
calibrated using near detector data, the beam MC, and cosmic ray data at
both detectors.  Once an analysis can correctly matches the real data
and the MC data, efficiencies and purities of the resulting sample can
be extracted from the MC truth information, systematic uncertainties
estimated, and expected sensitivity curves to the final physics
parameters calculated.  Only at this point is the ``box opened'', the
far detector data run through the analysis, and the hypotheses tested to
see what Mother Nature is really doing..

\subsection{Atmospheric sector neutrino oscillations}\label{sec:CC}

The main goal of the MINOS experiment is a precision measurement of the
$\nu_\mu$ disappearance oscillations first observed in atmospheric
neutrinos.  In the Standard Model, neutrinos are assumed to be massless
and direct neutrino mass measurements have been able to establish only
upper limits to their masses.  Quantum mechanics predicts that if
neutrinos do indeed possess a non-zero mass, then although the neutrinos
are created and interact via the weak force as flavor eigenstates
(corresponding to the flavors of leptons: electrons, muons and tauons --
$\nu_e,\nu_\mu,\nu_\tau$) they propagate through space as mass
eigenstates ($\nu_1,\nu_2,\nu_3$).  The flavor eigenstates are simple
superpositions of the mass eigenstates~\cite{pontecorvo}.  If the
neutrinos have differing masses, then the flavor of the neutrino varies
as these states drift into and out of phase with each other while
propagating through space, thus ``oscillating'' in flavor.  For the case
of two-flavor oscillations ({\it e.g.}
$\nu_\mu\leftrightarrow\nu_\tau$) the probability that a neutrino
produced via the weak interaction in the muon flavor state has
oscillated to, or will be detected as, the tau state by the time it
interacts is:
\begin{equation}
        P_{\nu_\mu\rightarrow\nu_\tau} = \sin^2 2\theta_{23}\sin^2\left (
        \frac{\Delta m^2_{32} L}{4 E_\nu}\right ),
\label{eq:probnumu-nutau}
\end{equation}
\noindent where the properties of nature being probed are the amplitude
or mixing angle $\theta_{23}$ and $\Delta m^2_{32} = m_3^2 - m_2^2$.
The observable quantities are the energy of the neutrino $E_\nu$ and the
distance the neutrino has traveled, also called the ``baseline'' $L$.
Observation of neutrino flavor oscillations which vary as $L/E$ implies
that both the terms $\Delta m^2_{32}$ and $\sin^22\theta_{23}$ are
non-zero, and that at least one of the participating neutrino flavors
has mass.

\begin{figure}[tb]
  \begin{center}
    \begin{minipage}[b]{0.45\textwidth}
      \includegraphics[width=\textwidth]{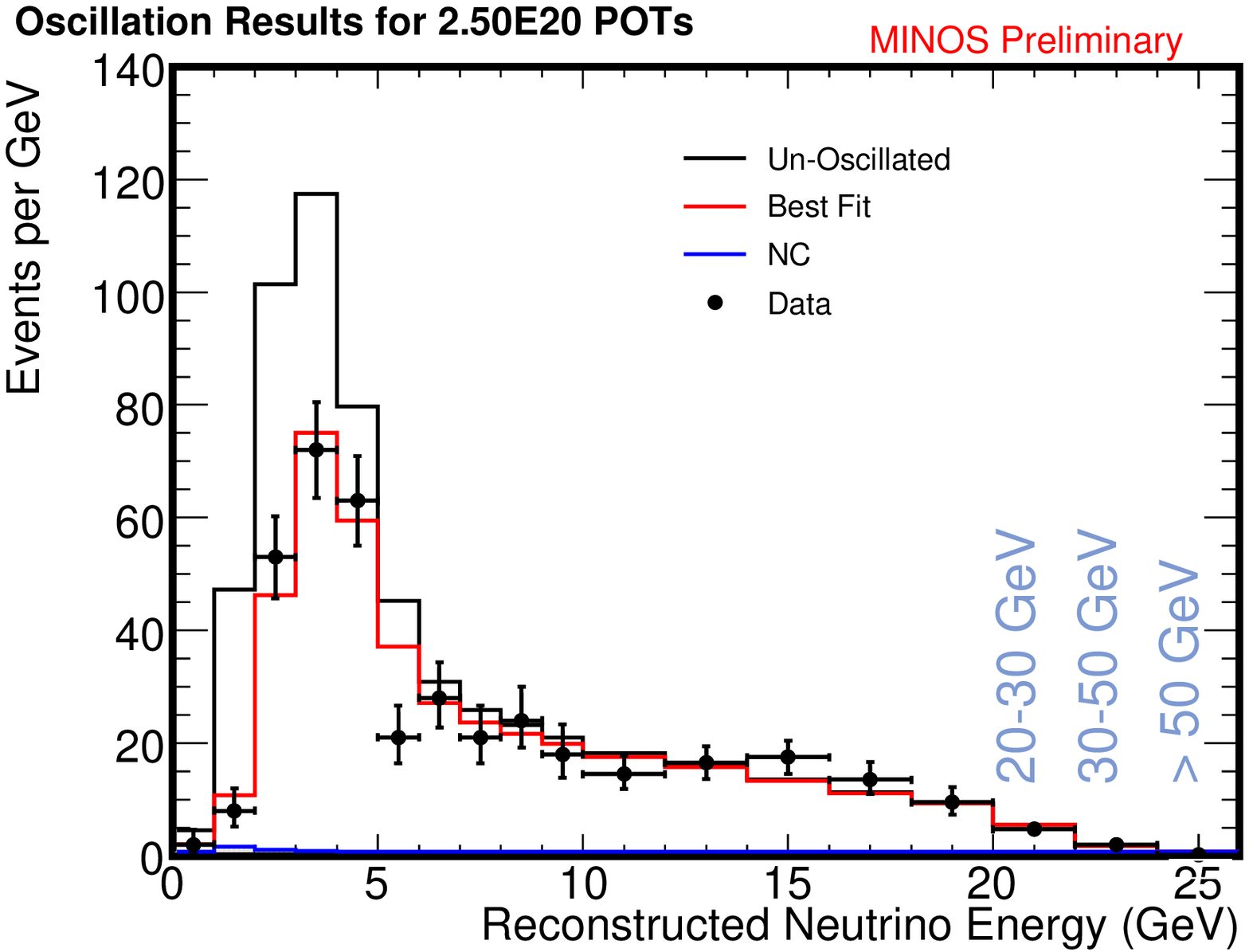}
    \end{minipage}
    \begin{minipage}[b]{0.45\textwidth}
      \includegraphics[width=\textwidth]{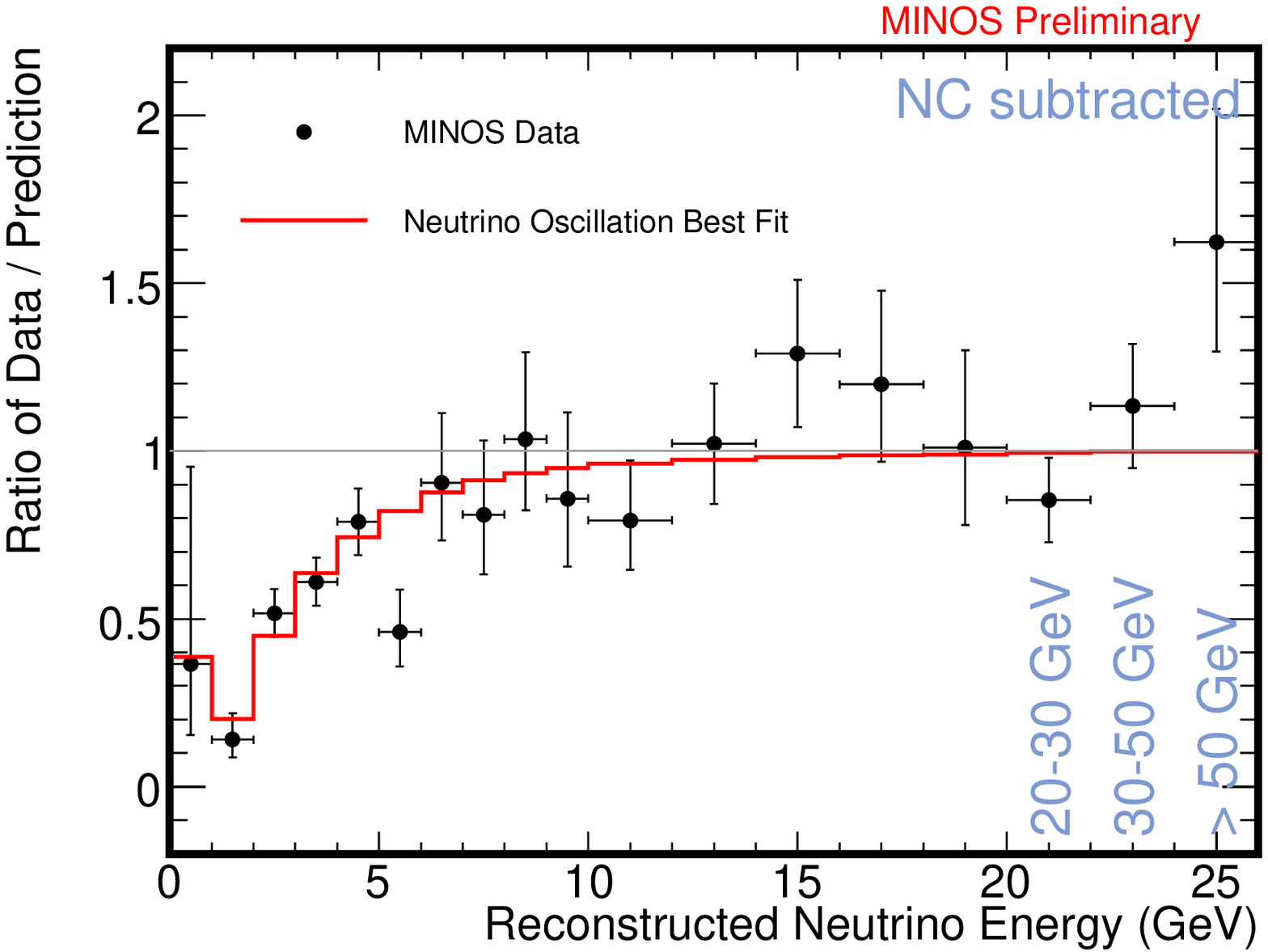}
    \end{minipage}
    \caption{(Left) The observed $\nu_\mu$ energy spectrum seen in the MINOS
      beam at the far detector for an exposure of 2.947$\times$10$^{20}$
      pot.  Black crosses are the data with statistical error
      bars, the black line the null hypothesis, the red line the
      expectations of the best fit oscillation scenario of $\left|\Delta
        m^2_{32}\right|=2.38^{+0.20}_{-0.16}\times\unit[10^{-3}]{eV^2}$,
      $\sin^22\theta_{23}=1.00_{-0.08}$,
      and the blue line (barely visible in the first few energy bins)
      the expected NC contamination.  (Right) The
      same quantities expressed as a ratio of observed over expected
      null hypothesis.}
    \label{fig:ccspectra}
  \end{center}
\end{figure}

The analysis techniques discussed above were applied to data from the
start of the NuMI beam through March~2007, totaling
2.947$\times$10$^{20}$ ``LE'' beam protons on target (``pot'').  This
includes the previously published\cite{minosprd} 1.27$\times$10$^{20}$
pot, although the analysis has been improved for both old and new data.
A 3\% larger fiducial volume was used, the data reconstruction was
improved and retained 4\% more good neutrinos, and the PID algorithm was
revamped to provide both better purity and efficiency.  The resulting
sample of 563~$\nu_\mu$ charged current (``CC'') neutrino interactions
is plotted as a function of reconstructed neutrino energy on the left of
Fig.~\ref{fig:ccspectra}, and a ratio with expectations (right) shows an
energy dependent deficit.

\begin{figure}[tb]
  \begin{center}
    \includegraphics[width=0.6\textwidth]{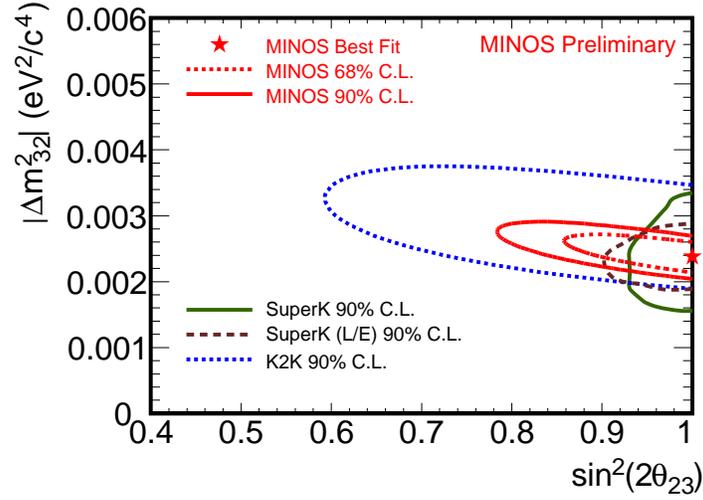}
    \caption{The allowed regions in the oscillation parameter space of
      Eq.~\ref{eq:probnumu-nutau}, obtained by fitting reweighted MC
      predictions to the MINOS data in Fig.~\ref{fig:ccspectra}.  MINOS
      results (red) at 68\% and 90\% c.l. are compared to Super-K
      results (green)\cite{superkprd,superkle} and K2K results (blue)
      \cite{k2k-prd} at 90\% c.l.}
    \label{fig:ccallowed}
  \end{center}
\end{figure}

Equation~\ref{eq:probnumu-nutau} was applied on a two-dimensional
($\Delta m^2, \sin^22\theta$) grid to the MC predictions, and a
$\chi^2$ formed compared to the data.  
Estimated systematic errors are 
less than the current statistical errors
and
applied as penalty terms to the $\chi^2$.  The best fit value for the
oscillation parameters to the MINOS data are $\left|\Delta 
  m^2_{32}\right|=2.38^{+0.20}_{-0.16}\times\unit[10^{-3}]{eV^2}$ and 
$\sin^22\theta_{23}=1.00_{-0.08}$, and the resulting 68\% and 90\%
confidence limit contours are shown in Fig.~\ref{fig:ccallowed}.


\subsection{Neutral Current Interactions}\label{sec:NC}

The $\nu_\mu$ disappearance results discussed in the previous section
(\ref{sec:CC}) use topological information to form a PID to select a
sample of CC $\nu_\mu$ neutrinos, on the assumption that the flavor they
are disappearing to is $\nu_\tau$, an active flavor of neutrino,
unobserved in MINOS since the bulk of the NuMI neutrino flux is at
energies below $\tau$ production threshold.  However, if the second
flavor of neutrino is a non-standard model sterile neutrino (one which
experiences no weak interactions), the disappearance signature could
look the same with very different underlying physics.

\begin{figure}[tb]
  \begin{center}
    \begin{minipage}[b]{0.45\textwidth}
      \includegraphics[width=\textwidth]{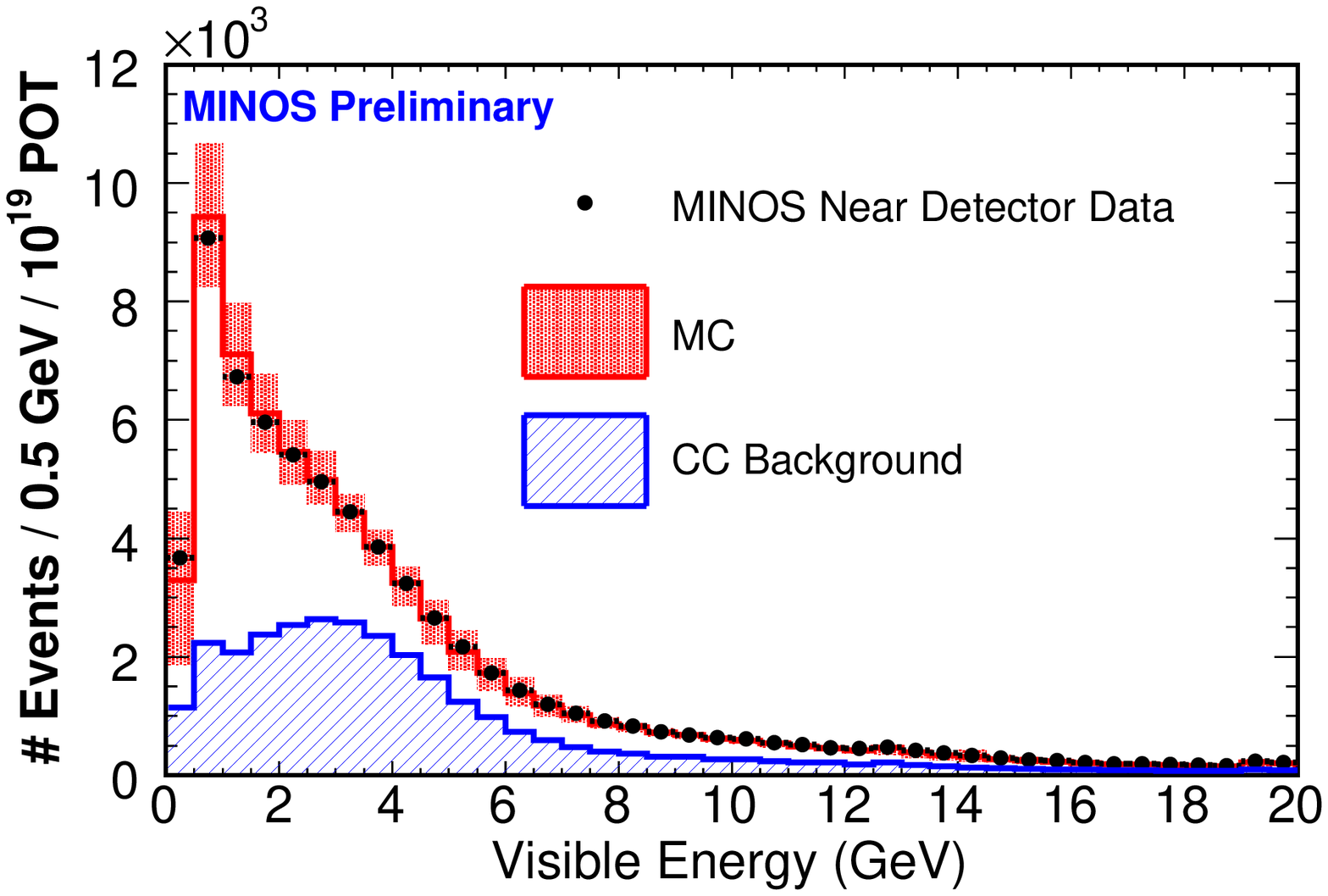}
    \end{minipage}
    \begin{minipage}[b]{0.45\textwidth}
      \includegraphics[width=\textwidth]{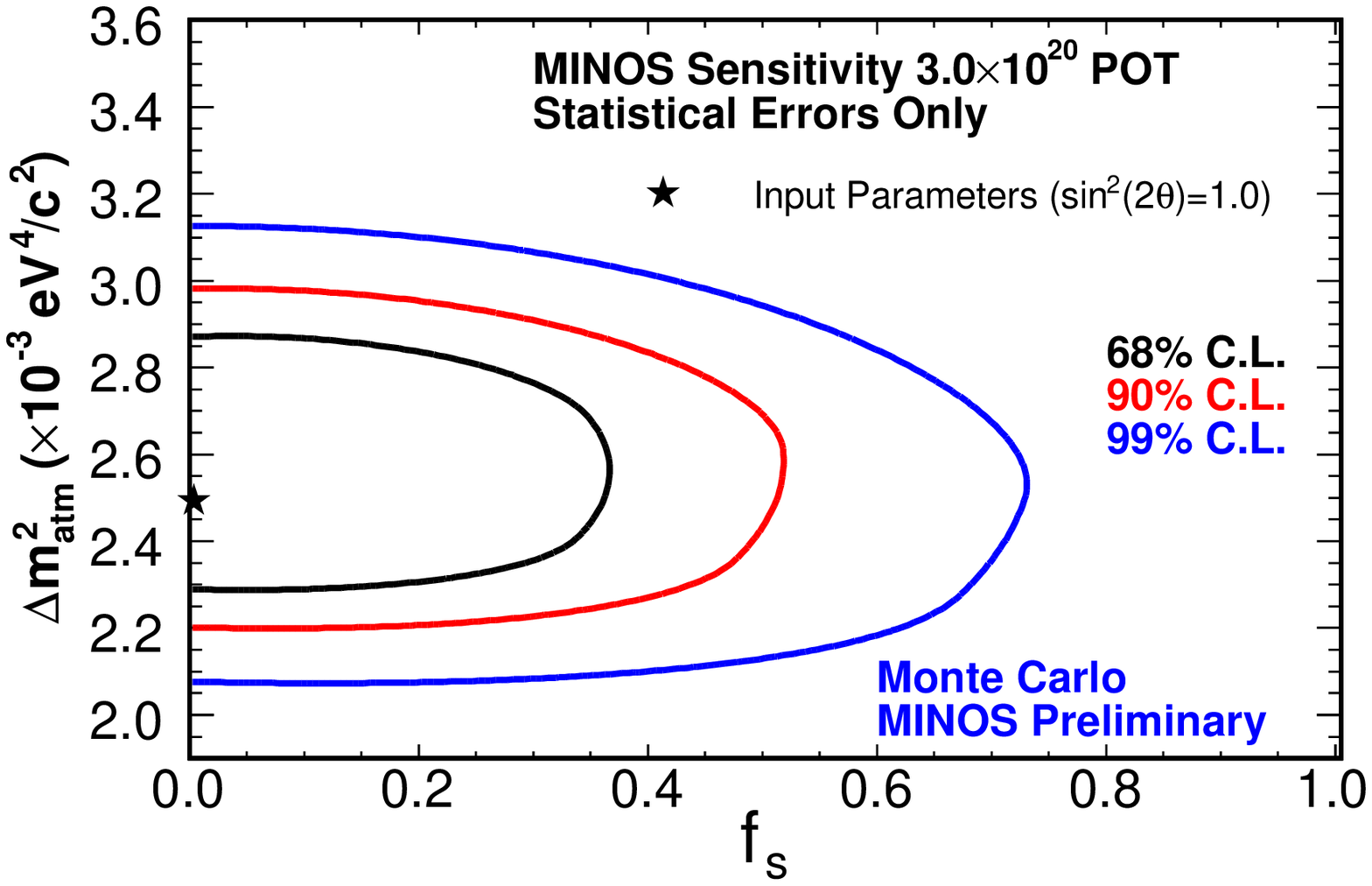}
    \end{minipage}
    \caption{(Left) The spectrum of NC-selected events seen in the near
      detector for an exposure of 3.0$\times$10$^{20}$ pot.
      The black dots are the reconstructed neutrino data, the red boxes
      the expected signal (systematic error bands as the area), and the
      blue region the expected CC contamination to this NC signal.
      (Right) The projected sensitivity to a fraction of $\nu_mu$
      disappearance to sterile rather than active neutrinos, as a
      function of $\Delta m^2$. }
    \label{fig:ncspectra}
  \end{center}
\end{figure}

NC neutrino interactions hold the key to separating these two scenarios
in MINOS.  Active neutrinos of any flavor can experience a NC $Z^0$
exchange with a nucleon in the detector and produce a diffuse
electromagnetic shower from the resulting $\pi^0$ decay to $\gamma
\gamma$.  A hypothetical sterile neutrino would not, so if some fraction
of the $\nu_\mu$ signal is changing to $\nu_s$, the NC spectrum would be
distorted and NC flux reduced.  A simple set of cuts has been applied to
the near detector data to select a NC-rich sample of neutrino
interactions for further study.  Short tracks ($<$60 planes) are
selected, then events with either no track at all or no track beyond
five planes from the shower are chosen.  The resulting spectrum of NC
near detector neutrino interactions is shown in the left of
Fig.~\ref{fig:ncspectra}, with projected limits on $f_s$ (the fraction
of disappearance to $\nu_s$ shown on the right if no NC disappearance is
observed.

This analysis is in the middle of the ``blind'' analysis scheme
discussed above.  Having chosen a set of topological cuts to select a NC
signal, data and MC comparisons are being made with near detector data
to verify that they are well understood before looking at the far
detector data to see what the potentially oscillated signal might look
like.  This analysis is expected to be complete the summer of 2008.

\subsection{Sensitivity to $\nu_e$ appearance }\label{sec:nue}

A third possibility for which particle the $\nu_\mu$'s are disappearing
to is $\nu_e$.  We know that there could be some natural mixing between
all three active flavors of neutrinos, and the amplitude of this is
parametrized as $\theta_{13}$.  The Chooz reactor experiment saw no
evidence of the converse $\nu_e$ disappearance at short baselines to
establish an upper limit on $\theta_{13}$\cite{chooz}.  However, the
presence of $\nu_e$ in the MINOS far detector beyond the low (1.3\%)
level inherent to the NuMI beam could provide evidence for a non-zero
$\theta_{13}$ below the Chooz limit, if the background of hadronic
showers masquerading as electromagnetic showers can be overcome.  The
PID algorithm used for $\nu_e$ selection uses a neural net technique to
pick out $\nu_e$-induced showers.  At the near detector the baseline is
far too short for $\nu_\mu$ to have oscillated to $\nu_e$, so any
observed $\nu_e$ must either be inherent in the beam or the
mis-reconstructed hadronic showers in question.  To improve the MC
estimations of what the levels of these backgrounds might really be in
the MINOS detectors, data-driven sensitivity studies have been performed
by close examination of near detector data tagged as $\nu_e$ events.

Two methods are used in these studies.  The first is to take the
well-understood class of CC events and subtract out those parts of the
event associated with the muon track, leaving only any hadronic
component near the interaction vertex caused by the nucleon's share of
the interaction energy.  These ``Muon-Removed Charged Current'' (MRCC)
events which are mis-classified as $\nu_e$ interactions are exactly the
sort of electromagnetic-dominated hadronic showers that form a large
part of the background for a $\nu_e$ appearance search.  There is a 20\%
discrepancy between the data and the MC predictions in both the standard
$\nu_e$ and MRCC samples with the MC overestimating the background.
Comparisons of standard data and MC shower topological distributions
disagree in the same way as does MRCC data with MRCC MC, confirming that
hadronic shower modeling is a major component of the disagreement.  The
MRCC sample is thus used to make and ad-hoc correction to the model to
NC events per bin, taking the beam  $\nu_e$ from the well-understood
beam MC.

A second method for estimating the $\nu_e$ background from hadronic
showers uses comparisons between the neutrino beam produced when the
focusing horn's current is turned off and the standard LE beam. The
actual composition of the selected $\nu_e$ events is quite different in
the two cases, allowing for the algebraic deconvolution of the different
background components by expressing the total number as a sum of the
different parts in the case of each beam:
\begin{equation}
  \begin{array}{rcl}
    N_{on} & = & N_{NC} + N_{CC} + N_e \\
    N_{off} & = & r_{NC}N_{NC} + r_{CC}N_{CC} + r_eN_e \\
  \end{array}
  \label{eq:nuehoo}
\end{equation}
\noindent where $N_{NC}$ and $N_{CC}$ are the numbers of background
events present originating from CC or NC interactions, $N_e$ the
inherent beam $\nu_e$ taken from the beam MC, and the $r$'s the ratios
that hold the differences between the two equations,
$r_{NC(CC,e)}=N_{NC(CC,e)}^{off}/N_{NC(CC,e)}^{on}$.  The horn on/off
ratios are extracted bin-by-bin in energy from the MC, are independent
of hadronic modeling, and match well between data and MC.  These
fractions can then be applied to the data itself to extract the
components of the background, indicating that there is 24\% too much CC
and 28\% too much NC backgrounds in the MC.  Checks with a third (pHE)
beam produce similar results, and both are compatible with the
corrections from the MRCC method outlined above.

\begin{figure}[tb]
  \begin{center}
    \includegraphics[width=0.6\textwidth]{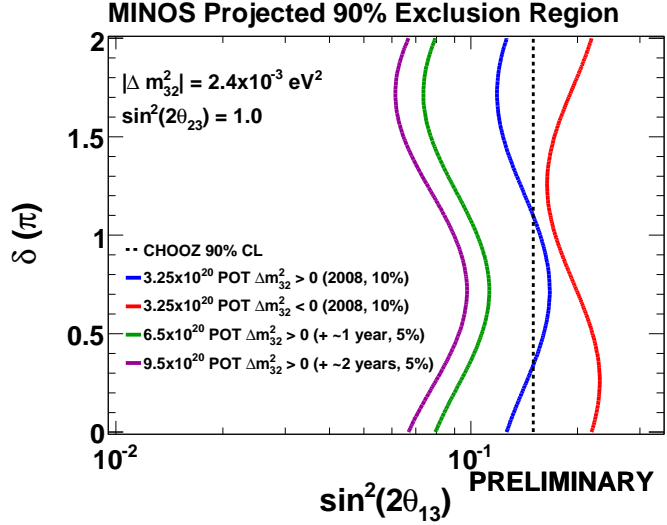}
    \caption{Projected 90\% c.l. limits on $\theta_{13}$ as a function
      of CP-violating $\delta$ from the MINOS experiment in the absence
      of a $\nu_e$ appearance signal.  These limits use data-driven
      background estimates from the near detector.  The vertical dashed
      line is the Chooz limit\cite{chooz}.  The rightmost (red) curve is
      the limit using the current exposure in the case of an inverted
      mass hierarchy, the neighboring (blue) curve is the limit from the
      same exposure if nature has a normal neutrino mass hierarchy.  The
      two curves on the left are the progressively more sensitive normal
      hierarchy limits for the increased NuMI beam exposure over the
      next two years.  The corresponding inverted hierarchy curves for
      these two scenarios are not shown for ease of viewing, but improve
      over the current exposure limits in a corresponding manner to the
      normal curves.}
    \label{fig:nue}
  \end{center}
\end{figure}

These data-drive backgrounds can then be extrapolated to the far
detector for use in establishing the sensitivity expected when using a
$\nu_e$ appearance search to try and measure $\theta_{13}$.  These
sensitivities are presented in Fig.~\ref{fig:nue} for three different
exposures, the current 3.25$\times$10$^{20}$ pot as well as those
expected for next two years.  The systematic errors for the current
background estimation are found to be 10\%, and with more data and study
it is projected to fall to 5\% for future years.  The unknown variable
of CP-violating $\delta$ contributes to $\nu_e$ appearance through the
matter effects on the beam between Fermilab and Soudan, so the y-axis of
these plots shows the effect of this $\delta$.  The actual sign of
$\Delta m^2{32}$ also enters in, making this analysis less sensitive for
the ``inverted'' mass hierarchy.  However, after two more years of
exposure MINOS will be sensitive to $\theta_{13}$ below the Chooz limit
for most combinations of $\delta$ and mass hierarchy.  The next step in
this blind analysis is to examine far detector data in ``sidebands''
that allow verification of techniques without being sensitive to actual
$\nu_e$ appearance. 

\section{Summary}\label{sec:summary}  

The MINOS long-baseline neutrino experiment has been receiving
\unit[735]{km} baseline neutrinos from the NuMI neutrino beam since
early 2005.  The primary experimental goal of a precision measurement of
the $\nu_\mu\leftrightarrow\nu_\tau$ disappearance oscillation
parameters has been achieved.  With the first 2.5$\times$10$^{20}$
protons on target, $\left|\Delta
  m^2_{32}\right|=2.38^{+0.20}_{-0.16}\times\unit[10^{-3}]{eV^2}$ and
$\sin^22\theta_{23}=1.00_{-0.08}$.  This is about a quarter of the
expected final exposure, which will allow fine distinction between
alternative disappearance hypotheses such as decoherence and neutrino
decay to be made in the future.  The first measurement of the spectrum
of neutral current neutrino interactions has been made in the
high-statistics near detector data.  When the blind analysis of the
corresponding far detector is complete later this year, it will be
sensitive to a sterile neutrino fraction $f_s\leq0.5$ at 90\%~c.l.
Again using the near detector data, a data-driven background estimate to
the $\nu_e$ appearance analysis has been made.  This yields a
sensitivity estimates comparable to the Chooz limit for the currently
available exposure of 3.25$\times$10$^20$ protons on target, reaching
several times lower than this limit as soon as next year.

The NuMI beam and the MINOS experiment are going strong, the data and
beam are well understood, and quality results are being produced.  The
next year should see the completion of initial analyses on all major
experimental goals and the continued refinement of the precision
parameter measurement of neutrino oscillations in the atmospheric
neutrino sector.

\section*{Acknowledgments}
We thank the Fermilab staff and the technical staffs of the
participating institutions for their vital contributions. This work was
supported by the U.S. Department of Energy, the U.S. National Science
Foundation, the U.K. Science and Technologies Facilities Council, and the
State and University of Minnesota.  
We gratefully
acknowledge the Minnesota Department of Natural Resources for their
assistance and for allowing us access to the facilities of the Soudan
Underground Mine State Park and the crew of the Soudan Underground
Physics laboratory for their tireless work in building and operating the
MINOS detector.

\end{document}